\documentclass{emulateapj}
\usepackage{amsmath}
\usepackage{natbib}
\usepackage{graphicx}
\usepackage{epstopdf}
\usepackage{relsize}
\usepackage[english]{babel}
\usepackage{ulem}
\usepackage[colorlinks,linkcolor={blue},citecolor={blue},urlcolor={blue}]{hyperref}
\usepackage{color}
\definecolor{revise}{rgb}{1, 0, 0}
\definecolor{grey}{rgb}{0.7, 0.7, 0.7}

\newcommand{\mpchi}{\,h^{-1}{\rm {Mpc}}}
\newcommand{\kms}{\,{\rm {km\, s^{-1}}}}
\newcommand{\msun}{\,h^{-1}{\rm M_{\sun}}}

\shorttitle{Galaxy Velocity Bias in Illustris} \shortauthors{J. Ye et al.}
\begin{document}
\title{Properties and Origin of Galaxy Velocity Bias in the Illustris Simulation}

\author{Jia-Ni Ye\altaffilmark{1,2},Hong Guo\altaffilmark{1},Zheng Zheng\altaffilmark{3}, \and Idit Zehavi\altaffilmark{4}}

\altaffiltext{1}{Key Laboratory for Research in Galaxies and Cosmology, Shanghai Astronomical Observatory, Shanghai 200030, China; guohong@shao.ac.cn}
\altaffiltext{2}{University of Chinese Academy of Sciences, 19 A Yuquanlu, Beijing 100049, China}
\altaffiltext{3}{Department of Physics and Astronomy, University of Utah, UT 84112, USA}
\altaffiltext{4}{Department of Astronomy, Case Western Reserve University, OH 44106, USA}

\begin{abstract}
We use the hydrodynamical galaxy formation simulations from the Illustris suite to study the origin and properties of galaxy velocity bias, i.e., the difference between the velocity distributions of galaxies and dark matter inside halos. We find that galaxy velocity bias is a decreasing function of the ratio of galaxy stellar mass to host halo mass. In general, central galaxies are not at rest with respect to dark matter halos or the core of halos, with a velocity dispersion above 0.04 times that of the dark matter. The central galaxy velocity bias is found to be mostly caused by the close interactions between the central and satellite galaxies. For satellite galaxies, the velocity bias is related to their dynamical and tidal evolution history after being accreted onto the host halos. It depends on the time after the accretion and their distances from the halo centers, with massive satellites generally moving more slowly than the dark matter. The results are in broad agreements with those inferred from modeling small-scale redshift-space galaxy clustering data, and the study can help improve models of redshift-space galaxy clustering.
\end{abstract}


\section{Introduction}
Galaxies are known to be biased tracers of the underlying dark matter in the universe \citep[see e.g.,][]{Kaiser84,Davis85,Bardeen86}. Although the spatial distribution of galaxies differs from that of the dark matter, we can still infer the dark matter distribution and constrain the cosmological parameters from galaxy clustering measurements by carefully taking into account the galaxy bias. A similar effect could also exist in the galaxy peculiar velocity field, i.e., galaxies may not exactly follow the dark matter velocity field, resulting in a `velocity bias'. The measurements of galaxy peculiar velocity field could essentially enhance the constraints on the linear growth rate through the correlation between galaxy velocity and density fields \citep[e.g.,][]{Kaiser87,Hamilton92,Percival09}. Observationally, direct measurements of the line-of-sight galaxy peculiar velocities require independent redshift-distance samples through Tully-Fisher relation, supernovae or Fundamental Plane, which are generally limited to the local universe and have significant measurement errors \citep[see e.g.,][]{Strauss95,Koda14}.

Although the galaxy peculiar velocity prevents us from measuring the real-space galaxy distribution, which is referred to as the redshift-space distortion effect, the encoded information, however, serves as another probe of the galaxy peculiar velocity field through measurements of galaxy clustering in redshift space. Large-scale galaxy redshift surveys, e.g., the Sloan Digital Sky Survey \citep[SDSS;][]{York00}, provide galaxy samples in a large volume that could possibly enable accurate statistical measurements of the galaxy peculiar velocity field on both linear and nonlinear scales. Theoretically, we can relate the galaxy peculiar velocity field to that of the dark matter in the halo modeling framework, i.e., by comparing the central galaxy velocity to that of the host halo and satellite galaxy velocities to those of the dark matter particles \citep[e.g.,][]{Berlind02,Tinker07}. Recently, \cite{Guo15a,Guo15c} measured and modeled the redshift-space galaxy clustering in the SDSS-II Main galaxy sample at $z{\sim}0.1$ \citep{Abazajian09} and SDSS-III Baryon Oscillation Spectroscopic Survey \citep[BOSS;][]{Eisenstein11,Dawson13} at $z{\sim}0.5$. They found that there is non-vanishing galaxy velocity bias for both central and satellite galaxies of different luminosities at different redshifts, which means that central galaxies are not at rest with respect to the halo centers of mass (see also \citealt{Reid14}) and the satellite galaxies generally do not move coherently with the dark matter velocity field. 

Satellite galaxy velocity bias has been found in hydrodynamical simulations for host halos of different masses \citep[see e.g.,][]{Yoshikawa03,Berlind03,Wu13b}. It naturally arises when the satellite galaxies experience the complex physical processes (such as the dynamical friction, tidal heating/stripping and mergers) after accretion (see Section 4 of \citealt{Guo15a} for more discussion). Mild central galaxy velocity bias with respect to the halo center of mass is also detected in previous low-resolution hydrodynamical simulations \citep{Yoshikawa03,Berlind03}. \cite{Bosch05} found a similar central galaxy velocity bias by comparing the velocities of central galaxies to the mean motion of satellite galaxies in the galaxy group catalogs of Two-Degree Field Galaxy Redshift Survey (2dFGRS) and the SDSS \citep[see also][]{Skibba11}. 

The baryonic effects on galaxy and dark matter distributions have been studied extensively in literature. For example, \cite{Jing06} and \cite{Rudd08} found that the matter power spectrum would be significantly affected by the baryonic processes at halo scales. \cite{Daalen14} predicted more than $10\%$ errors on galaxy-galaxy and galaxy-mass clustering on sub-Mpc scales based on collision-less models that neglect baryonic effects. \citet{Weinberg08} found that good agreements of halo occupation statistics between subhalos in a dark-matter only simulation and galaxies in a hydrodynamic simulation, while subhalos are more easily depleted in the densest regions of the most massive halos. \cite{Velliscig14} claimed that the physical processes relevant to galaxy formation can change halo properties (e.g., mass, mass profile, and mass function) over four orders of magnitude in mass. \cite{Bocquet16} found that at low redshifts the halo number densities in hydrodynamical simulations would decrease by up to $15\%$ compared to dark-matter only simulations for masses less than $10^{14} M_{\odot}$.

While the baryonic effects on the spatial distribution and halo mass might be more important, the velocity field, however, is less affected. In a recent work with a suite of hydrodynamical simulations from the Evolution and Assembly of GaLaxies and their Environments (EAGLE) project \citep{Schaye15}, \cite{Hellwing16} found that the existence of baryons only causes variations of less than 3\% for dark matter velocity power spectrum on scales of $1 h\,{\rm Mpc}^{-1} \leq k \leq 10 h\,{\rm Mpc}^{-1}$. They also reported negligible baryonic influence on the peculiar velocities of halos with masses larger than $3\times10^{11}\msun$ (the change is at the level of $1\kms$), which lends support to the theoretical modeling of galaxy velocity bias based on dark matter only simulations as in \cite{Guo15a}.

The measurements of galaxy velocity bias have important applications in various studies. For example, \cite{Farahi16} found that systematic uncertainties in the satellite velocity bias measurements would dominate the error budget of the estimated halo masses for galaxy clusters, which is also confirmed by \cite{Sifon16}. As investigated by \cite{Reid14}, galaxy velocity bias needs to be taken into account for precise constraints (at a level 2.5\%) on the growth rate of cosmic struture. However, the origin of central galaxy velocity bias, i.e., the relative motion between central galaxies and halo centers, is still not very clear.

As proposed in \cite{Bosch05}, there are two potential physical scenarios to explain central galaxy velocity bias, either the central galaxy is not relaxed in a virialized halo (referred to as the non-relaxed galaxy scenario) or the central galaxy is relaxed at the halo potential minimum while the halo itself is not relaxed (referred to as the non-relaxed halo scenario). In reality, both of the two scenarios may play a role in the measured velocity bias, as the mergers between central and satellite galaxies and those between different halos could both affect the velocity bias. Given the importance of the constraining power of galaxy peculiar velocity field in cosmology, we will use state-of-art high-resolution hydrodynamical simulations in this paper to understand the origin of galaxy velocity bias, and its dependence on various galaxy and halo parameters, which could provide better parameterization and more accurate modeling of galaxy velocity bias and aid the interpretation of measurements in the real universe. 

This paper is organized as follows. Section~\ref{sec:data} provides the details of the hydrodynamical simulations used and our methodology in measuring galaxy velocity bias. We presents the results in Section~\ref{sec:results} and conclude in section~\ref{sec:conclusion}. 

\section{Simulation and Method} \label{sec:data}
\subsection{The Illustris Simulation} \label{subsec:sim}
\setcounter{footnote}{0}
In this paper, we use the hydrodynamical simulations from the Illustris suite\footnote{publicly available at http://www.illustris-project.org/data/} \citep{Genel14,Vogelsberger14a,Vogelsberger14b}, which were run with the moving-mesh code \texttt{AREPO} \citep{Springel10}. The set of Illustris simulations consists of six full volume runs in a $(75\mpchi)^3$ box, with three resolution levels of hydrodynamical simulations and the corresponding dark-matter only $N$-body simulations \citep{Nelson15}. The `full' baryonic physics model in the hydrodynamical simulations traces the co-evolution of five different types of elements, including gas cells, dark matter particles, passive gas tracers, stars and stellar wind particles, and supermassive black holes from redshift $z=127$ to $z=0$ \citep{Vogelsberger14b}. The galaxy formation model includes the physical processes of gas cooling and photo-ionization, star formation and interstellar medium model, stellar evolution and feedback, black holes and supermassive black hole feedback. These simulations are shown to reproduce many of the observed galaxy properties, such as the galaxy luminosity function and the cosmic star formation rate density \citep{Vogelsberger14a}. The cosmological parameters in the Illustris suite are $\Omega_m=0.2726$, $\Omega_\Lambda=0.7274$, $\Omega_b=0.0456$, $\sigma_8=0.809$, $n_s=0.963$, and $h=0.704$, consistent with the WMAP-9 measurements \citep{Hinshaw13}.

We focus on the highest-resolution run (Illustris-1) at $z=0$ and the corresponding dark matter only (DMO) run with the same initial conditions (Illustris-1-Dark). The mass resolutions for the dark matter and baryon particles in Illustris-1 are $6.26\times10^6\,\rm{M_\sun}$ and $1.26\times10^6\,\rm{M_\sun}$, respectively. The mass resolution for the dark matter particles in Illustris-1-Dark is $7.52\times10^6\,\rm{M_\sun}$, which is the sum of the dark matter and baryon components in Illustris-1. The dark matter halos in the simulations are identified using the friends-of-friends (FOF) algorithm \citep{Davis85} with a minimum of 32 particles. The subhalos are then identified in the FOF groups using the \texttt{SUBFIND} algorithm \citep{Springel01,Dolag09}. Other types of particles in Illustris-1 are associated with each FOF group in a secondary linking stage \citep{Dolag09}. To study the evolution history of the galaxy velocity field, we also use the subhalo merger trees in the Illustris suite which are constructed using the \texttt{SubLink} algorithm \citep{Rodriguez-Gomez15}. The cross-matching between the full physics and the DMO runs is made available in Illustris-1 by finding the corresponding subhalos with the largest number of matching dark matter particles in the two simulations, as the total number of dark matter particles in the two simulations are the same. We can therefore investigate the effect of the baryon physics on the galaxy velocity field. 

In the \texttt{SUBFIND} algorithm, the largest subhalo in each FOF group is defined as the \textit{main halo} or \textit{central subhalo} hosting the central galaxy, and all other subhalos in the group are regarded as the substructure that hosts the satellite galaxies \citep{Springel01}. Since the mass of the central subhalo does not include those in other subhalos of the same FOF group, it is different from the definitions in \cite{Guo15a,Guo15c} where the halo mass is defined to include the substructure. To be consistent with \cite{Guo15a,Guo15c} and to avoid confusion, in what follows, the term `\textit{subhalo}' can be used for the central subhalo or the substructure, while we will only use the terms `\textit{halo}' or `\textit{host halo}' to refer to the group that includes all subhalos in the FOF group (i.e., central subhalos and other substructure). The mass of each halo, $M_{\rm h}$, is calculated by adding the mass of all its subhalos. It is, however, different from the mass of the whole FOF group because there is still a non-negligible fraction of fuzz particles in the FOF group that are not gravitationally bound to any subhalos. To make fair comparisons with the results from the Illustris-1-Dark simulation, the halo or subhalo mass in Illustris-1 should be calculated as the total mass including all particle types. 

The halo center is defined to be the same as that of the central subhalo, i.e., the spatial position of the most bound particle (which can be of any particle type in Illustris-1) in the central subhalo. In our fiducial model, we define the halo velocity, $\mathbf{v}_{\rm h}$, as the average velocity of the dark matter particles in the halo, i.e., the center-of-mass velocity for dark matter particles. This is different from the definition of the FOF group velocity available in the Illustris-1 catalog, which is computed as the sum of the mass weighted velocities of {\it all} particle types. But it does not make a big difference as the dark matter dominates the mass contributions in the halos. We will discuss the effects of the different halo velocity definitions in the following sections.

A central (satellite) galaxy in Illustris-1 is defined by the stellar particles or cells (excluding the wind particles) enclosed within twice the stellar half-mass radius around the center of each halo (subhalo). So the central and satellite galaxies take the same positions as their hosting halos and subhalos, respectively. The galaxy velocity is computed as the mass weighted sum of the stellar velocities in each galaxy, because different stellar cells may have different masses. We impose a low stellar mass limit for the galaxy sample (both central and satellite galaxies) used in this paper as $10^9\msun$, which corresponds to about $1,500$ stellar particles and ensures secure measurements of the galaxy properties. After checking in the lower resolution hydrodynamical simulations in the Illustris suite, we find that the results in this paper do not vary with the simulation resolution once the low-mass limit of the galaxy sample includes at least $1,000$ stellar particles. The original subhalo catalog in Illustris-1 at $z=0$ includes $4,366,546$ individual subhalos, while our final galaxy sample after applying the mass limit has only $22,548$ galaxies with $15,075$ of them being centrals. There are 59 satellite galaxies having a surrounding subhalo with less than 32 dark matter particles, so we can treat them as `orphan' galaxies. Moreover, $2,242$ galaxies ($19$ centrals and $2,223$ satellites) in Illustris-1 do not have counterparts in the DMO run, which means that about $0.1\%$ of the halos and $30\%$ subhalos would be disrupted without the existence of the baryons.     

\subsection{Galaxy Velocity Bias Measurements}
In \cite{Guo15a,Guo15c}, the galaxy velocity bias is modeled separately for the central and satellite galaxies in the halo occupation distribution (HOD) framework \citep{Zheng16}. The central galaxy velocity bias, $\alpha_{\rm c}$, is expressed in the distribution of the non-zero 3D velocity offset, $\Delta\mathbf{v}=\mathbf{v}_{\rm c}-\mathbf{v}_{\rm h}$, between the central galaxies and the host halos. The probability distribution of each component of $\Delta\mathbf{v}$, $\Delta v_i$, is assumed to be a Laplace distribution in \cite{Guo15c} as, 
\begin{equation}
f(\Delta v_i)=\frac{1}{\sqrt{2}\sigma_{\rm c}}\exp\left(-\frac{\sqrt{2}|\Delta v_i|}{\sigma_{\rm c}}\right),
\end{equation}
which is consistent with what we find in the Illustris simulation. The standard deviation of the distribution, $\sigma_{\rm c}$, is parameterized as the central velocity bias $\alpha_{\rm c}$ times the 3D dark matter particle velocity dispersion $\sigma_{\rm v}$ in the host halo, $\sigma_c = \alpha_c \sigma_{\rm v}$. The dark matter velocity dispersion is computed as $\sigma_{\rm v}^2=\langle||\mathbf{v}_{\rm p}-\mathbf{v}_{\rm h}||^2\rangle$, where $\mathbf{v}_{\rm p}$ is the 3D velocity of each dark matter particle. When there is no central galaxy velocity bias, $\alpha_{\rm c}=0$ and the central galaxy velocity exactly follows that of the host halo. In \cite{Guo15c}, the velocity bias is measured in terms of the line-of-sight velocity offset, while in the simulations we are able to measure the 3D velocity offset to better characterize the velocity bias.

The spatial positions of the satellite galaxies in the HOD model are represented by randomly selected dark matter particles in the halos, and their relative velocity to the halo center is scaled by a satellite velocity bias, $\alpha_{\rm s}$, as
\begin{equation}
\mathbf{v}_{\rm s}-\mathbf{v}_{\rm h} = \alpha_{\rm s} (\mathbf{v}_{\rm p}-\mathbf{v}_{\rm h}),
\end{equation}
where $(\mathbf{v}_{\rm p}-\mathbf{v}_{\rm h})$ is the relative velocity of the selected dark matter particle to the halo center. Therefore, the velocity dispersion of the satellite galaxies within the halos, $\sigma_{\rm s}$, is expressed as $\sigma_{\rm s}=\alpha_{\rm s}\sigma_{\rm v}$. The situation without satellite galaxy velocity bias corresponds to $\alpha_{\rm s}=1$. 

Since in the hydrodynamical simulations we can directly measure the peculiar velocities for both central and satellite galaxies, the galaxy velocity bias can be computed for a given sample of central or satellite galaxies using the same formula as,
\begin{equation}
	\alpha_{\rm *}=\sqrt{\langle\alpha_{\rm *}^2\rangle}=\sqrt{\left<\frac{||\mathbf{v_{\rm *}}-\mathbf{v_{\rm h}}||^2}{\sigma_{\rm v}^2}\right>}, \label{eq:velbias}
\end{equation}
where the subscript `*' can be `c' or `s' for central and satellite galaxies, respectively. Equation~\ref{eq:velbias} is consistent with the definitions in \cite{Guo15c} and is an unbiased estimator of the galaxy velocity bias. Throughout the paper, the measurement errors on the galaxy velocity bias are estimated with the bootstrap resampling method. 

\section{Results} \label{sec:results}

In this section, we present our main findings for the velocity bias in the Illustris simulations. We study the dependence of velocity bias on a variety of halo properties, including halo mass, reference frame of halo velocity, halo formation time, and halo density profile. In each case, the dependence on the galaxy property (in terms of the stellar-to-halo mass ratio) is shown. We also present an explicit example with a halo of mass $M_h=10^{12}\msun$ to show the relation between the velocity evolution of central and satellite galaxies and the merging/growth histories of the galaxies and the host halo.

\subsection{Dependence on Mass}
\begin{figure*}
	\epsscale{1.0}
	\plotone{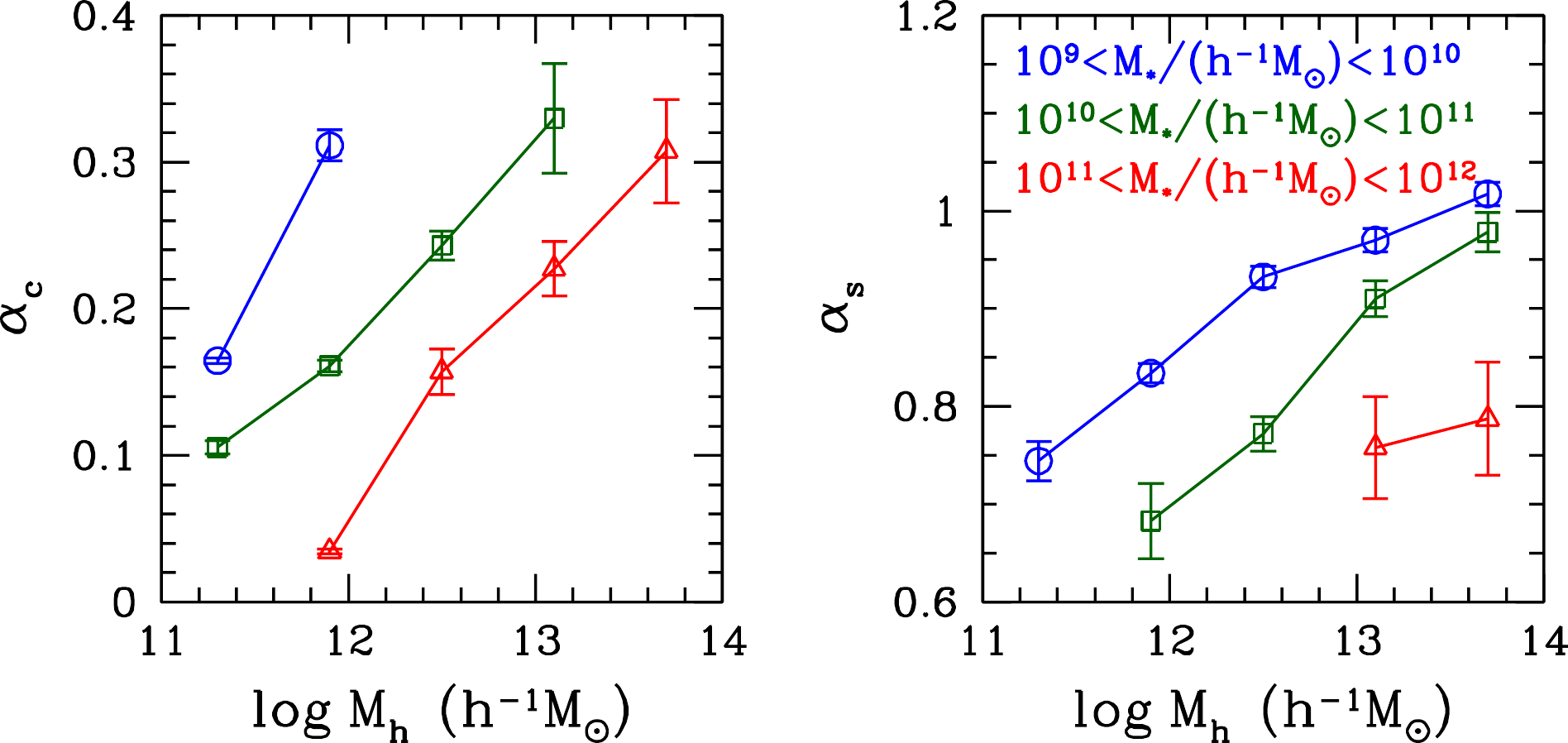}
	\caption{Dependence of central (left panel) and satellite (right panel) galaxy velocity bias on the host halo mass for galaxies of different stellar masses. The blue circles, green squares and red triangles are for the galaxy stellar mass $M_*$ in the range of $10^9$--$10^{10}\msun$, $10^{10}$--$10^{11}\msun$, $10^{11}$--$10^{12}\msun$, respectively.} \label{fig:hostmass}
\end{figure*}
We show in Figure~\ref{fig:hostmass} the dependence of the central (left panel) and satellite (right panel) galaxy velocity bias on the host halo mass for different stellar mass samples as labeled. Both central and satellite galaxy velocity biases show strong dependence on the host halo mass, with more massive halos having larger $\alpha_{\rm c}$ and $\alpha_{\rm s}$. There is also strong dependence on the galaxy stellar mass. For halos of the same mass, more massive galaxies have smaller velocity bias values. The stellar and halo mass dependences imply that the galaxy velocity bias could be affected by interaction between galaxies and their host halos. 

The halo mass dependence of central galaxy velocity bias tends to favor the scenario that more massive halos are less relaxed as they were generally formed more recently. The central velocity bias will be affected by the change in the halo velocity due to halo mergers. But the dependence on the central galaxy stellar mass indicates that the mass growth history of the central galaxies is also an important factor. For satellite galaxies, the stellar mass dependence is consistent with the expectation that dynamical friction in the halo environment will slow down the motion of the satellite galaxies and the effect is stronger for more massive satellites \citep{Chandrasekhar42,Binney08}. Overall, the halo mass dependence of galaxy velocity bias can be explained by the earlier formation time of lower mass halos for central galaxies and by the dynamical friction effect for satellite galaxies.

\begin{figure*}
	\epsscale{1.0}
	\plotone{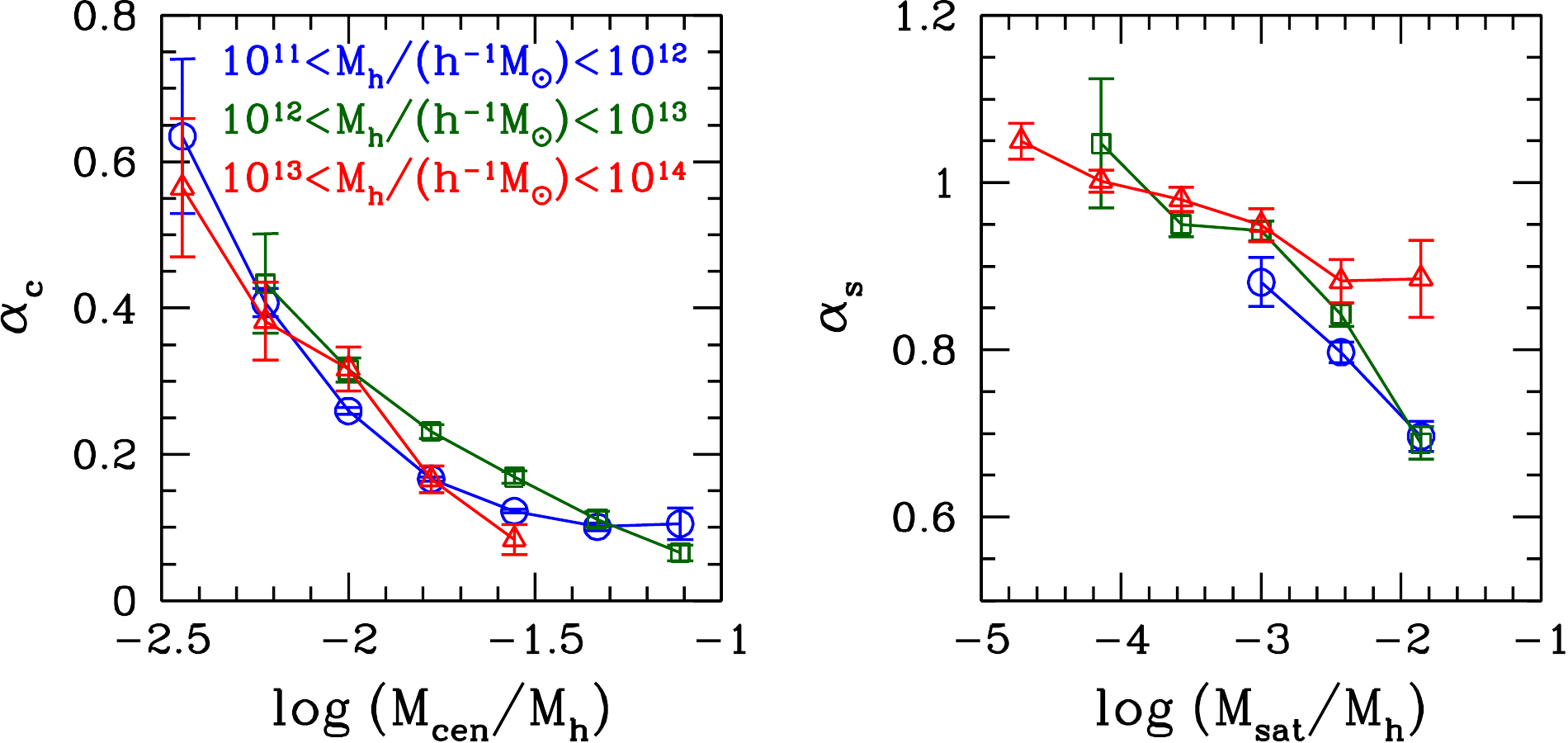}
	\caption{Relation between the galaxy velocity bias and the SHMR in different halo mass bins for central (left panel) and satellite (right panel) galaxies. The blue circles, green squares and red triangles are for the halo mass bins of $10^{11}$--$10^{12}\msun$, $10^{12}$--$10^{13}\msun$, and $10^{13}$--$10^{14}\msun$, respectively.}
	\label{fig:massratio}
\end{figure*}
There is a trend in Figure~\ref{fig:hostmass} that galaxies with similar stellar-to-halo mass ratios (SHMRs) tend to have similar velocity bias for both central and satellite galaxies. We check the correlation between the galaxy velocity bias and the SHMR for different halo masses in Figure~\ref{fig:massratio}. Both the central and satellite velocity biases are decreasing with SHMR, and there is no strong dependence on the halo mass. We also verify that for a given SHMR, the dependence of galaxy velocity bias on the stellar mass is also very weak. It is clear that the stellar and halo mass dependences shown in Figure~\ref{fig:hostmass} mostly come from the dependence on the SHMR.

\subsection{Dependence on Halo Velocity Definition}\label{subsec:veldef}
\begin{figure*}
	\epsscale{1.0}
	\plotone{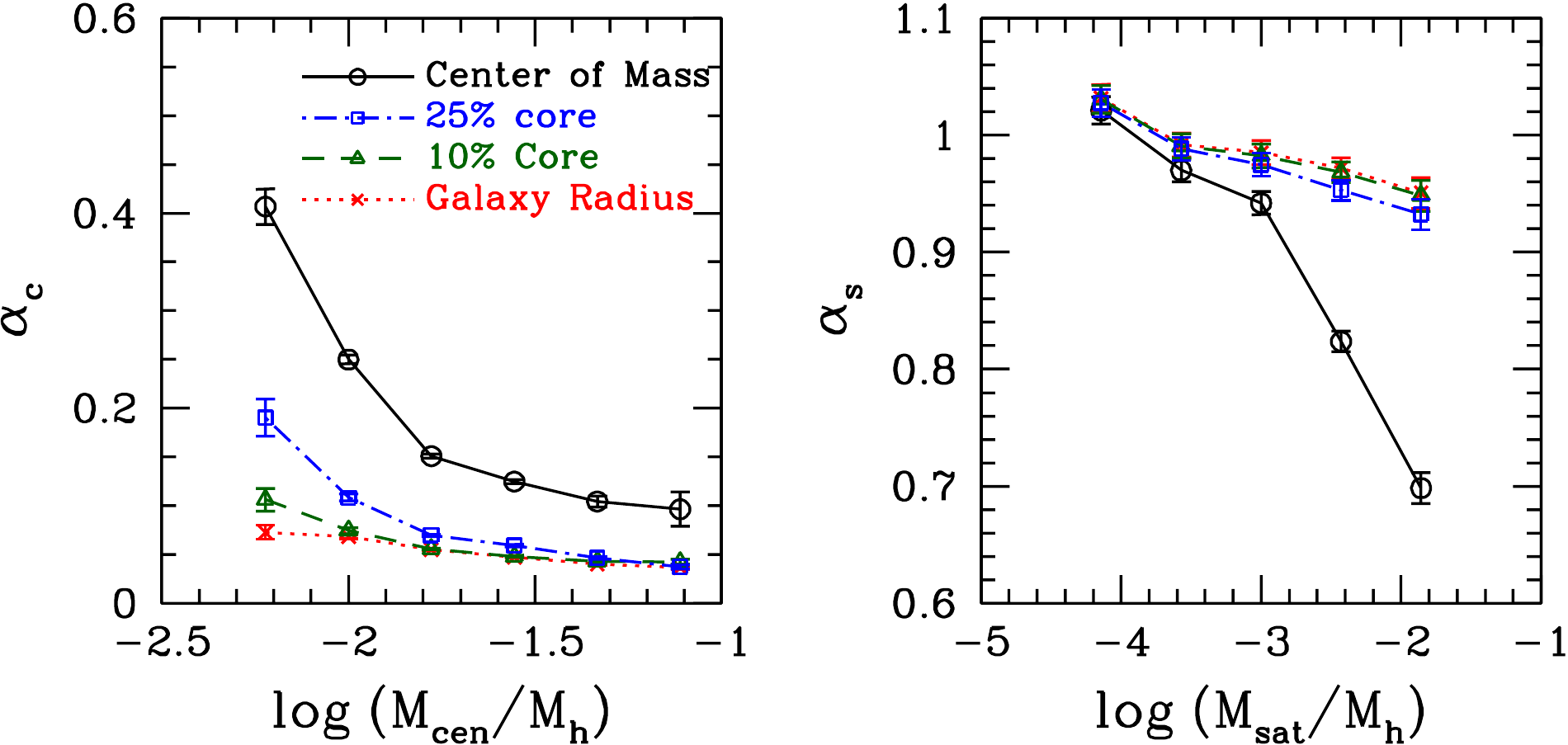}
	\caption{Measurements of the galaxy velocity bias using different halo velocity definitions of central (left panel) and satellite (right panel) galaxies for all galaxies in halos of different masses in the sample. The solid, dot-dashed, short-dashed, and dotted lines with symbols are for the different halo velocity definitions as labeled. We consider the average velocities of all dark matter particles (center-of-mass velocity) as our fiducial model. The average velocities of the innermost $25\%$ and $10\%$ of the dark matter particles, and those within twice the stellar half-mass radius of the central galaxies are also shown for comparison.}
	\label{fig:core}
\end{figure*}
The galaxy velocity bias would potentially be affected by the merger events between halos due to the changes in the halo bulk velocities. Meanwhile, as the average velocity of dark matter particles depends on radius \citep[]{Behroozi13}, galaxy velocity bias will then be dependent on the halo velocity definition \citep{Reid14,Guo15a}. Our fiducial model defines the halo velocity as the average velocity of all dark matter particles in each halo (center-of-mass velocity), but there are also other definitions commonly used in the literature. For example, the halo velocity in the \texttt{ROCKSTAR} halo finder \citep{Behroozi13} is defined as the average velocity of the dark matter particles within the innermost $10\%$ of the halo radius, which would be less affected by the disturbance in the outer part of the halo during merger events. \cite{Guo15a} define the halo velocity as the bulk velocity of the inner 25\% of the particles \citep[as also used in][]{Li12}. To investigate the effect of the different halo velocity definitions, we compare the effect on the galaxy velocity bias between our fiducial model and three other halo velocity definitions, the average core velocities of the innermost $25\%$ and $10\%$ of the dark matter particles, and also those within the central galaxy radius (twice the stellar half-mass radius). 

The equation for computing galaxy velocity bias should be changed accordingly for different halo velocity definitions. The central velocity bias can still be calculated using Equation~\ref{eq:velbias}. However, for satellite galaxies, when the halo core velocity $\mathbf{v}_{\rm core}$ is used instead of the halo bulk velocity $\mathbf{v}_{\rm h}$, the denominator in Equation~\ref{eq:velbias} should be replaced by $\sigma_{\rm v,core}$, i.e., $\sigma_{\rm v,core}^2=\langle||\mathbf{v}_{\rm p}-\mathbf{v}_{\rm core}||^2\rangle$ \citep{Guo15a}. We note that the intent of $\sigma_{\rm v,core}$ is not the velocity dispersion of the dark matter particles in the halo core velocity frame, but rather to ensure that whether $\alpha_{\rm s}$ equals to 1 does not depend on the halo velocity definition. So when we detect a satellite velocity bias signal for one halo velocity definition, it will also exist in other definitions.

Figure~\ref{fig:core} compares the galaxy velocity bias parameters under different halo velocity definitions for the dependence on the SHMR. For central galaxies, the dependence of $\alpha_{\rm c}$ on the SHMR becomes much weaker for the halo velocity definitions using dark matter particles enclosed within a smaller radius around the center. The value of $\alpha_{\rm c}$ increases with the adopted radius of enclosed dark matter particles, which means that the central galaxy follows closely the local dark matter velocity field rather than the dark matter bulk velocity of the whole halo. It agrees with the paradigm that while the halo bulk velocity can be significantly changed during mergers, the central part of the halo is less affected. However, even for the dark matter enclosed within the galaxy radius, the velocity bias $\alpha_{\rm c}$ is not zero. The central velocity bias measurements of different halo velocity definitions tend to have a lower limit around $\alpha_{\rm c}{\sim}0.04$ for different SHMRs. 

The trend for the satellite galaxies is similar. The samples with the halo velocity definitions using dark matter particles in smaller radii have larger values of $\alpha_{\rm s}$ and closer to unity for $M_{\rm sat}/M_{\rm h}>10^{-4}$ (i.e., less biased). But the dependence on the SHMR is still very significant in all cases. As in the situation of the central galaxies, the satellite velocity bias is generally non-vanishing for any of the halo velocity definitions. Defining the average dark matter velocity within the galaxy radius is not practical for theoretical modeling, and different values of $\sigma_{\rm v,core}$ need to be used to infer velocity bias for galaxies of different stellar masses (sizes). Based on the above results, the velocity bias for central galaxies defined in such a way is minimal. 

\subsection{Dependence on Formation Time}
\begin{figure*}
	\epsscale{1.0}
	\plotone{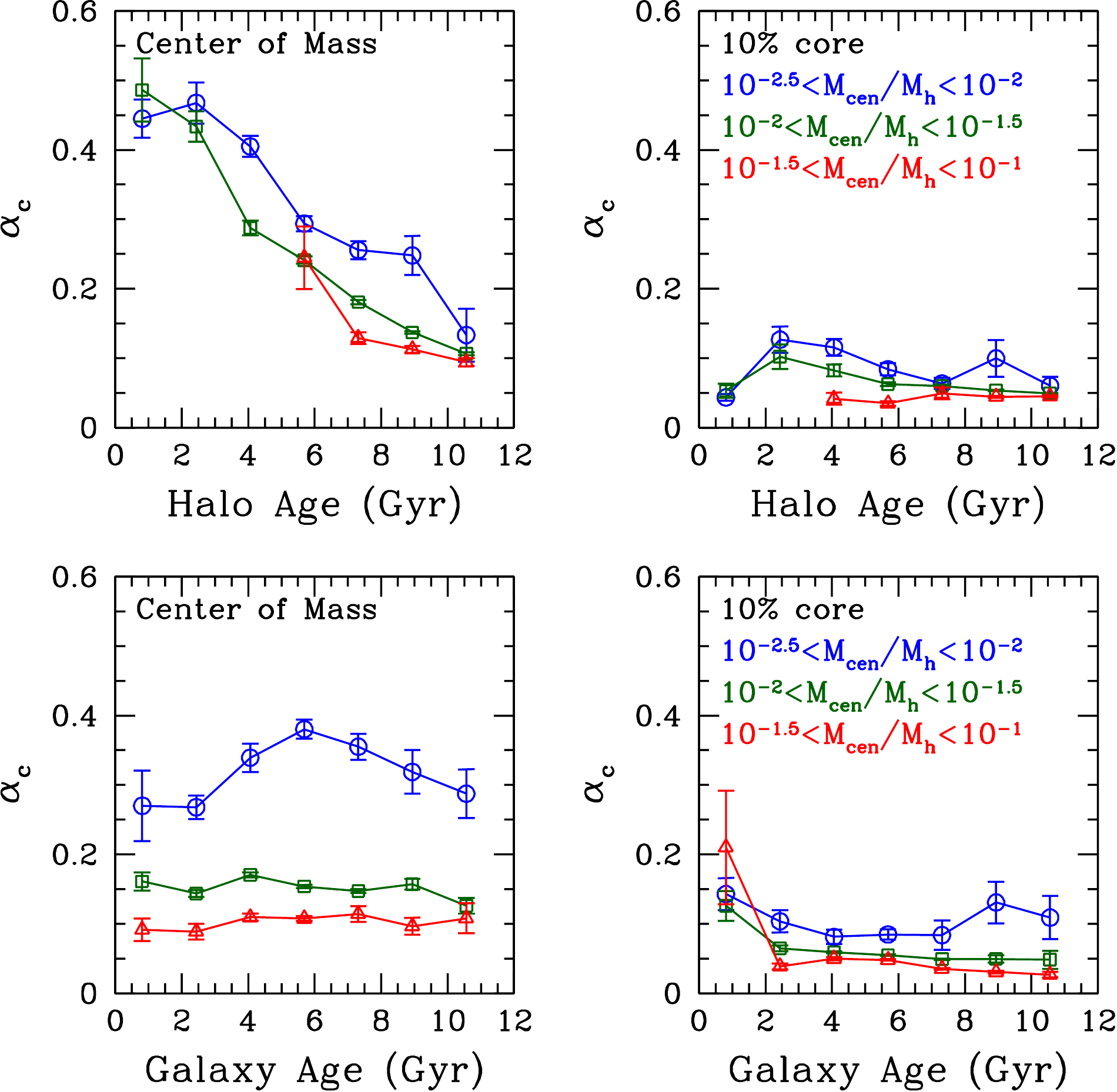}
	\caption{Dependence of the central galaxy velocity bias on the halo (top panels) and galaxy (bottom panels) ages for different SHMRs in different colors, as labeled. The left panels show our fiducial model of the center-of-mass velocity frame and the right panels display the result for the halo velocity definition with $10\%$ core (see Section~\ref{subsec:veldef}).
}
	\label{fig:ac_zform}
\end{figure*}
As the galaxy velocity bias can be affected by merger events of halos and galaxies, we expect the galaxy velocity bias to also depend on the halo formation time, because more massive halos form later and statistically undergo more mergers. We define the halo formation time as the redshift when the host halo assembled half of its final mass at $z=0$, which can be done by tracing the main progenitor branch of the halo merger tree at all the output snapshots \citep[see details in][]{Rodriguez-Gomez15}. Because the satellite galaxies suffer from mass loss after they are accreted onto the host halos as a result of the intra-halo effects (e.g., tidal stripping), we define the formation time of a galaxy as the redshift when its stellar mass reaches half of the peak stellar mass over the whole merger history. Changing the formation time definitions to other ones only has minimal effects on our results. 

\begin{figure*}
	\epsscale{1.0}
	\plotone{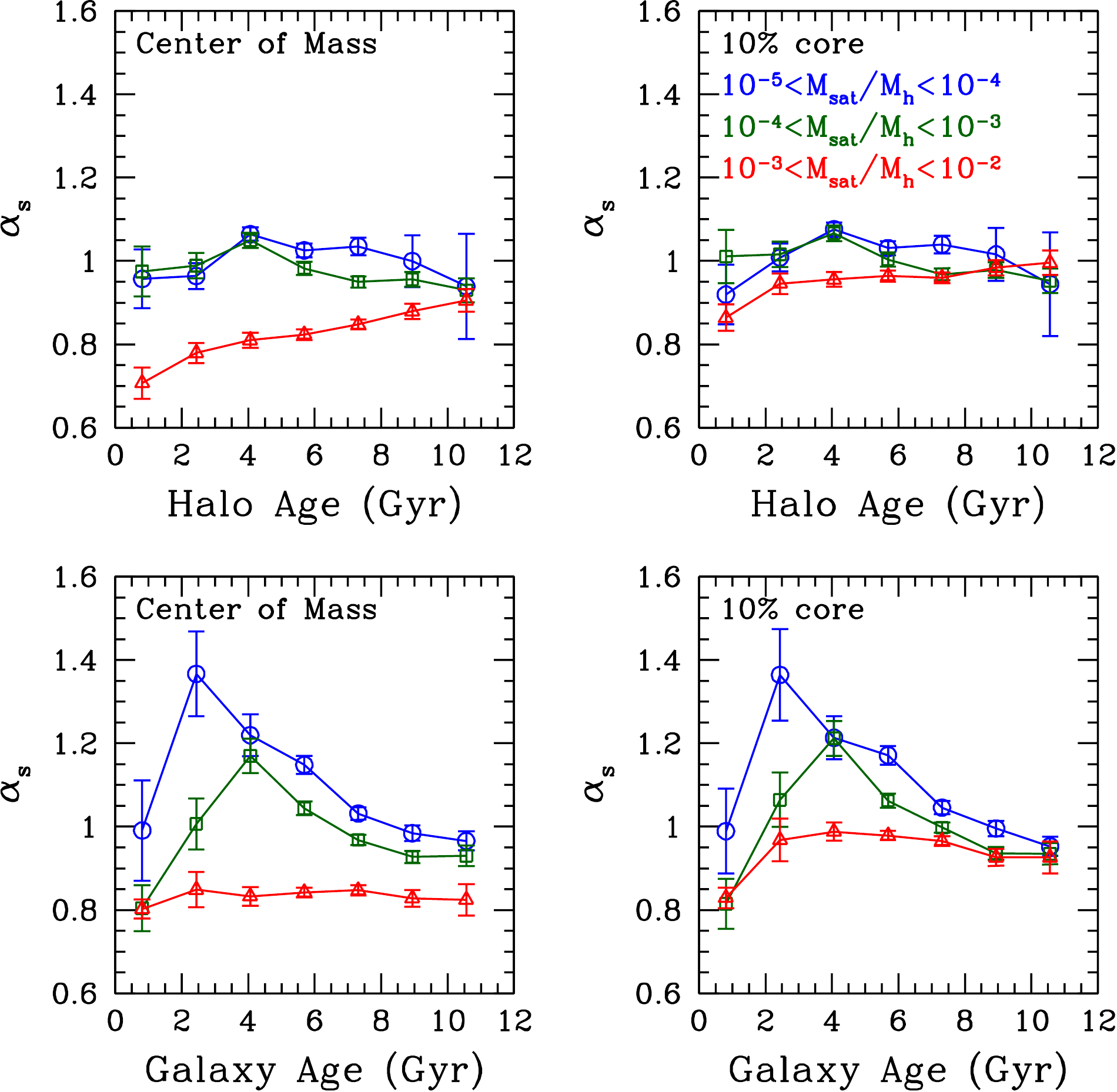}
	\caption{Similar to Figure~\ref{fig:ac_zform}, but for the satellite galaxy velocity bias.}
	\label{fig:as_zform}
\end{figure*}

We show in Figures~\ref{fig:ac_zform} and~\ref{fig:as_zform} the dependences on the halo and galaxy ages for central and satellite galaxies, respectively. The dependence is displayed for samples of different SHMRs and we show both our fiducial model of the center-of-mass velocity frame (left panels) and the result with the halo velocity using the $10\%$ core (right panels), as defined in the previous section. In the center-of-mass halo velocity frame, the central velocity bias strongly depends on the halo age, with older halos having smaller $\alpha_{\rm c}$. There is almost no dependence on the galaxy age for $M_{\rm cen}/M_{\rm h}>10^{-2}$. For central galaxies with $M_{\rm cen}/M_{\rm h}<10^{-2}$, most of them reside in low-mass halos of ${\sim}10^{11}\msun$. The peak of $\alpha_{\rm c}$ around galaxy ages of about $6$~Gyrs is related to the dependence on the halo age, because many of the halos hosting these galaxies are formed more recently. Although the halo and galaxy ages are generally correlated with each other with older halos hosting older galaxies, the large scatter between the two makes the dependence on the galaxy age very weak. When we switch to the halo velocity frame using the average dark matter velocity within the $10\%$ core, the dependence of $\alpha_{\rm c}$ on halo age becomes much weaker and it also does not show any strong dependence on the galaxy age.  

The age dependence for the satellite galaxy velocity bias is more complicated. For samples of $M_{\rm sat}/M_{\rm h}<10^{-3}$, $\alpha_{\rm s}$ has weak dependence on the halo age but strongly depends on the galaxy age. The situation is reversed for satellite galaxies of $M_{\rm sat}/M_{\rm h}>10^{-3}$. This may be caused by a combination of accretion onto host halos, tidal stellar mass loss, and dynamical evolution of satellite galaxies in halos. Changing the halo velocity definitions does not significantly affect the dependences on the halo and galaxy ages. 

\begin{figure*}
	\epsscale{1}
	\plotone{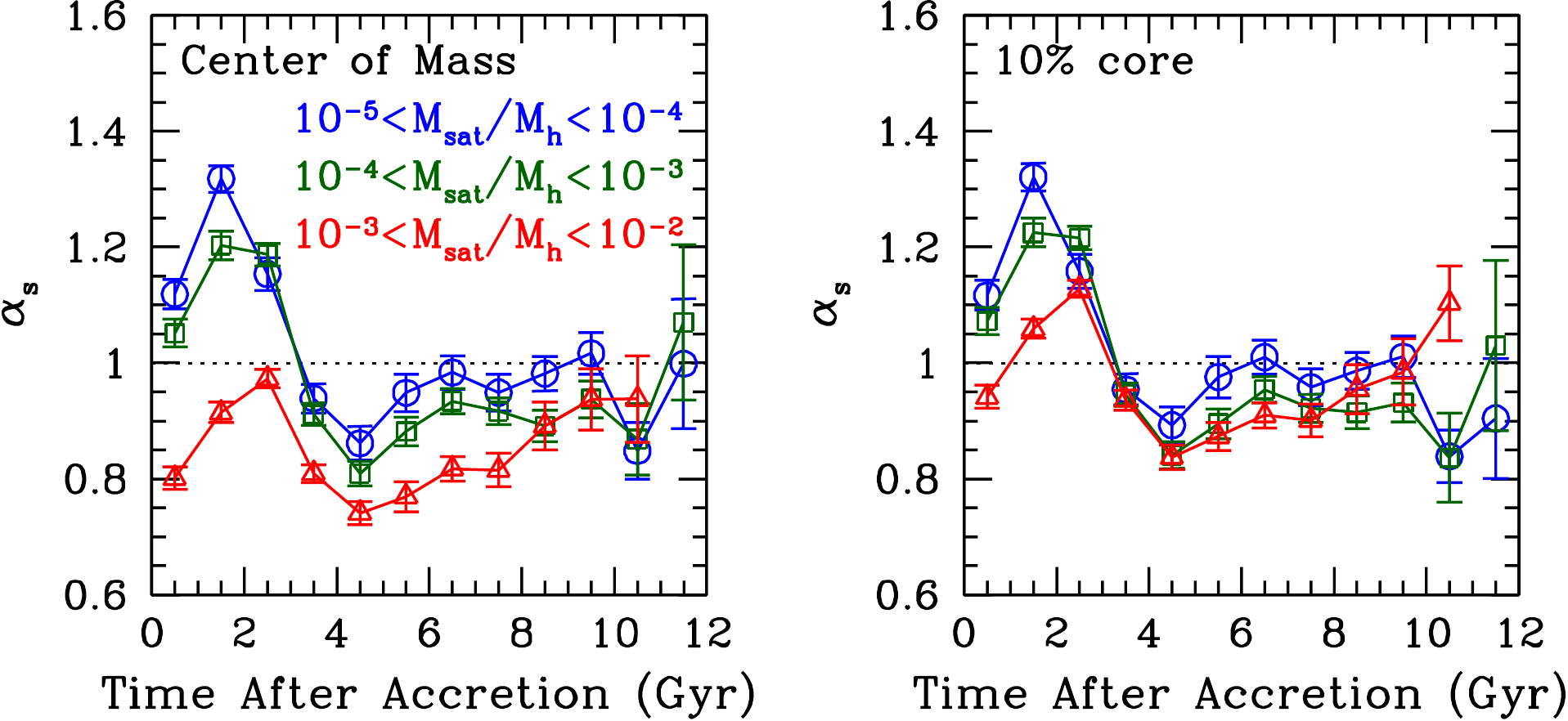}
	\caption{Dependence of satellite velocity bias on the time after the satellite galaxies were accreted onto the halos for the center-of-mass halo velocity (left) and $10\%$ core velocity (right) frames. Lines with different symbols are for satellite galaxies of different SHMRs as labeled.}
	\label{fig:as_infall}
\end{figure*}
\begin{figure*}
	\epsscale{1}
	\plotone{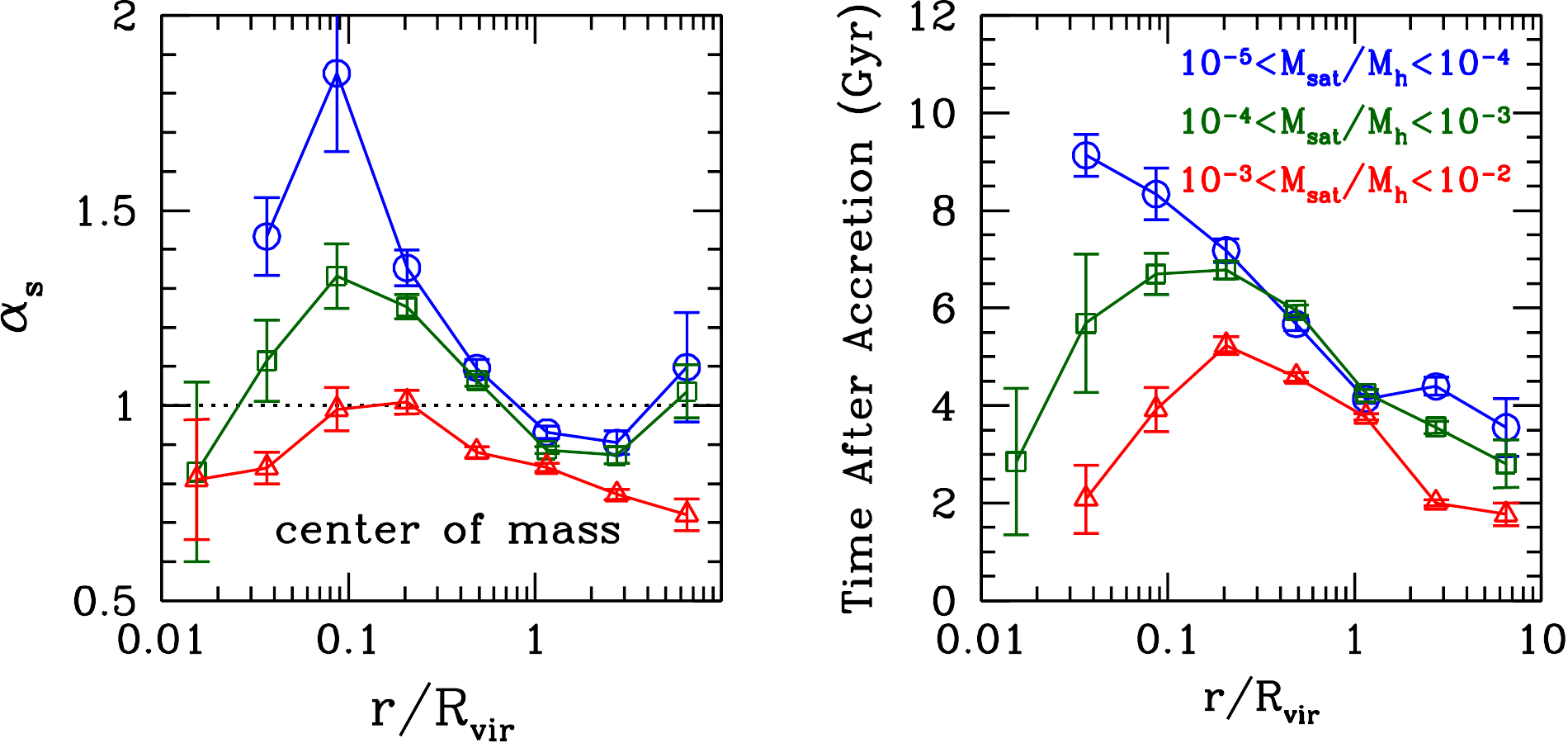}
	\caption{Left: Dependence of $\alpha_{\rm s}$ on the distances of the satellite galaxies to the halo centers for the samples of different SHMRs. Right: Average time after accretion for satellite galaxies at different radii in the halos. The lines with different symbols are for samples of different SHMRs.
}
	\label{fig:as_radius}
\end{figure*}
As satellite galaxies in the halo environment will undergo the physical processes such as tidal stripping/heating, galaxy harassment and dynamical friction, the value of satellite velocity bias is then potentially related to the time after they were accreted onto the halos. We show in Figure~\ref{fig:as_infall} the dependence of $\alpha_{\rm s}$ on the time after accretion for the center-of-mass halo velocity (left panel) and $10\%$ core velocity (right panel) frames. The strong and complex dependence on the time after accretion reflects the satellite dynamics in the halos. For satellite galaxies with low SHMRs, when they were accreted onto the halos, they would roughly follow the dark matter velocity field. The satellite galaxies are then accelerated towards the halo potential minimum and pass the halo center in about 2~Gyrs, which is consistent with the merger timescale \citep{Jiang08}. They are further decelerated due to dynamical friction and then start the relaxation process, which will take another few Gyrs. 

Under such a paradigm, we expect the satellite galaxy velocity bias to be also dependent on the distance to the halo center, as shown in the left panel of Figure~\ref{fig:as_radius}. \cite{Berlind03} also investigated the relation of the satellite galaxy velocity bias as a function of radius (their Figure~17), but did not find any significant dependence on radius. The mass resolution of the baryonic particles in their smoothed particle hydrodynamics (SPH) simulation is $8.5\times10^8\,M_{\odot}$, which will affect the identification of the low-mass satellite galaxies. 

The radius dependence for the satellite galaxies with large SHMRs is weaker than those with smaller SHMRs. The satellite velocity bias decreases with radius for distance of $r<0.1\,R_{\rm vir}$, where $R_{\rm vir}$ is the virial radius of the host halo and is defined as twice the radius encompassing half of the halo mass. Such a decreasing trend may be caused by the strong tidal stripping effects around the halo centers. The satellite galaxies with large SHMRs generally have lower values of $\alpha_{\rm s}$, compared to other samples. As shown in Figure~\ref{fig:massratio}, $\alpha_{\rm s}$ shows a sharp decrease with the SHMR for $M_{\rm sat}/M_{\rm h}>10^{-3}$. It is related to the fact that the dynamical friction is more efficient for these massive satellites in the halo environment. 

As shown in the right panel of Figure~\ref{fig:as_radius}, the discrepancies in the dependence of $\alpha_{\rm s}$ on radius for different SHMR samples is related to their different evolution histories. For samples with $M_{\rm sat}/M_{\rm h}>10^{-3}$, the inner regions of the halos ($r/R_{\rm vir}<0.1$) are dominated by recently accreted (less than 4~Gyrs) satellite galaxies, while the satellite galaxies accreted at earlier epochs would have higher chances to reach equilibrium in the halos and populate the regions with higher radii around $r/R_{\rm vir}{\sim}0.5$. For samples with smaller SHMRs, satellite galaxies that were accreted earlier will preferentially occupy the inner part of the halos as a consequence of dynamical friction.

\subsection{Dependence on Density Profile}
\begin{figure*}
	\epsscale{1.0}
	\plotone{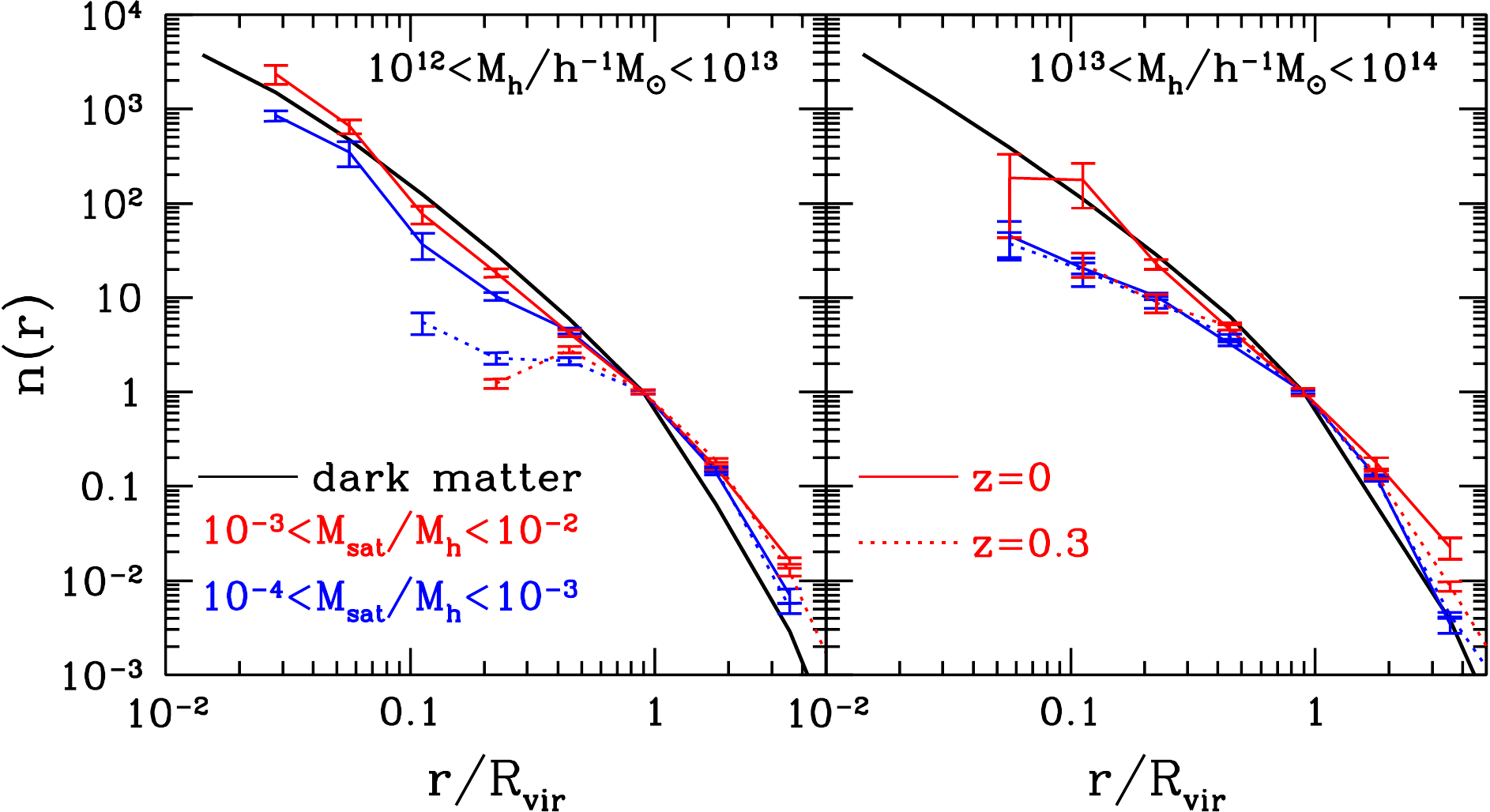}
	\caption{Normalized density profiles of the satellite galaxies in halos of different masses and for different SHMRs. The density profiles are normalized at the virial radius $R_{\rm vir}$ for better comparison. The black solid line in each panel is the average density profile for the dark matter distributions in these halos. The dotted lines are for the satellite density profile at the epoch of $z=0.3$, while the solid lines are measured at $z=0$.
}\label{fig:profile}
\end{figure*}
By modeling the observed galaxy clustering of the SDSS-III BOSS galaxy sample in the HOD framework, \cite{Guo15a} found that there exists degeneracy between the value of $\alpha_{\rm s}$ and the satellite density distribution profile within halos, in the direction of lower values of $\alpha_{\rm s}$ for steeper satellite density profiles (their Figure~11). 

In Illustris-1, we can directly measure the satellite galaxy density profiles and compare with the dark matter distributions. We show in Figure~\ref{fig:profile} the normalized density profiles of the satellite galaxies in halos of different masses and for different SHMRs (solid lines with different colors). The density profiles are normalized at the virial radius $R_{\rm vir}$ for better comparisons. The black line in each panel is the average density profile for the dark matter distributions in these halos. 

We find that the satellite galaxies with higher SHMRs have steeper density slopes and show better agreement with the dark matter spatial distribution in the halos. This is in line with the steeper density slope for subhalos of higher masses in a host halo \citep[see e.g., Figure~1 of][]{Guo16}. The satellite galaxy distribution is generally steeper than that of the subhalos because of the existence of the `orphan galaxies', without surrounding subhalos \citep[see, e.g., Figure~2 in][]{Vogelsberger14b}. But for the galaxy samples considered in this paper (with stellar mass larger than $10^9\msun$), the orphan galaxies only occupy about $0.7\%$ of the satellite galaxies. Thus, they would not have any significant effect on the satellite density profiles. 

The satellite galaxies with $M_{\rm sat}/M_{\rm h}>10^{-3}$ have consistent spatial distributions with the dark matter particles, while the spatial profile of the satellites with a lower SHMR is much shallower than that of the dark matter in massive halos of $M_{\rm h}>10^{13}\msun$, as shown in the right panel of Figure~\ref{fig:profile}. For satellites with $10^{-4}<M_{\rm sat}/M_{\rm h}<10^{-3}$, the density slope is slightly steeper in less massive halos in the left panel of Figure~\ref{fig:profile} than those in the right panel, which is consistent with the weak mass dependence of $\alpha_{\rm s}$ for satellites of the same SHMR shown in Figure~\ref{fig:massratio}. 

However, according to Figure~\ref{fig:massratio}, the average value of $\alpha_{\rm s}$ for satellite galaxies with $10^{-3}<M_{\rm sat}/M_{\rm h}<10^{-2}$ is about $0.8$ (center-of-mass frame), i.e. $\sigma_{\rm s}=0.8\sigma_{\rm v}$. These satellite galaxies are moving at a speed about $0.8$ times that of the dark matter particles in the halo, although their spatial distributions are consistent with those of the dark matter. It seems inconsistent at first sight, since the velocity bias between the satellite galaxies and the dark matter will make the satellite galaxy distributions in the next snapshot deviate from those of the dark matter. 

We show in Figure~\ref{fig:profile} the corresponding satellite galaxy density profiles at redshift $z=0.3$ for different SHMRs as the dotted lines. This is done by tracing the same set of satellite galaxies at $z=0$ back to $z=0.3$, and only selecting galaxies that were satellites at $z=0.3$. For satellite galaxies with $10^{-3}<M_{\rm sat}/M_{\rm h}<10^{-2}$ (measured at $z=0$), their density slopes are much shallower at $z=0.3$ than at $z=0.0$ for different host halo masses, which means that the satellite galaxies at higher redshifts are more likely to distribute in the outer part of the halo. According to Figure~\ref{fig:as_infall}, after about 4~Gyrs of the accretion epoch, the satellite velocity bias for galaxies with this SHMR is increasing with time and closer to unity. Although the density slope at higher redshifts is much shallower than the dark matter distribution, the satellite galaxies will gradually move toward the halo centers because they are moving slower than the dark matter particles. The density slope is therefore gradually increased to be similar to that of the dark matter.

The situation is similar for satellite galaxies with $10^{-4}<M_{\rm sat}/M_{\rm h}<10^{-3}$ in halos of $10^{12}$--$10^{13}\msun$, where the density slope is also increased from $z=0.3$ to $z=0$. However, for the corresponding satellite galaxies in halos of $10^{13}$--$10^{14}\msun$, the difference between the density slopes at the two redshifts is tiny. It may be related to the fact that the value of $\alpha_{\rm s}$ for the satellite galaxies in these massive halos is closer to unity. When the satellite galaxies have similar velocity distribution profile as the dark matter, their density distribution would evolve very slowly, in spite of the much shallower density slope compared to that of the dark matter.

The overall behavior of the satellite velocity bias and distribution profile, in comparision to dark matter, indicates that the phase space distribution of
satellite galaxies has not reached equilibrium and the system is not in a steady state. This implies that one should be cautious when applying the Jeans 
equations to study such systems.

\subsection{An Explicit Example of the Velocity Bias}
\begin{figure*}
	\epsscale{1.0}
	\plotone{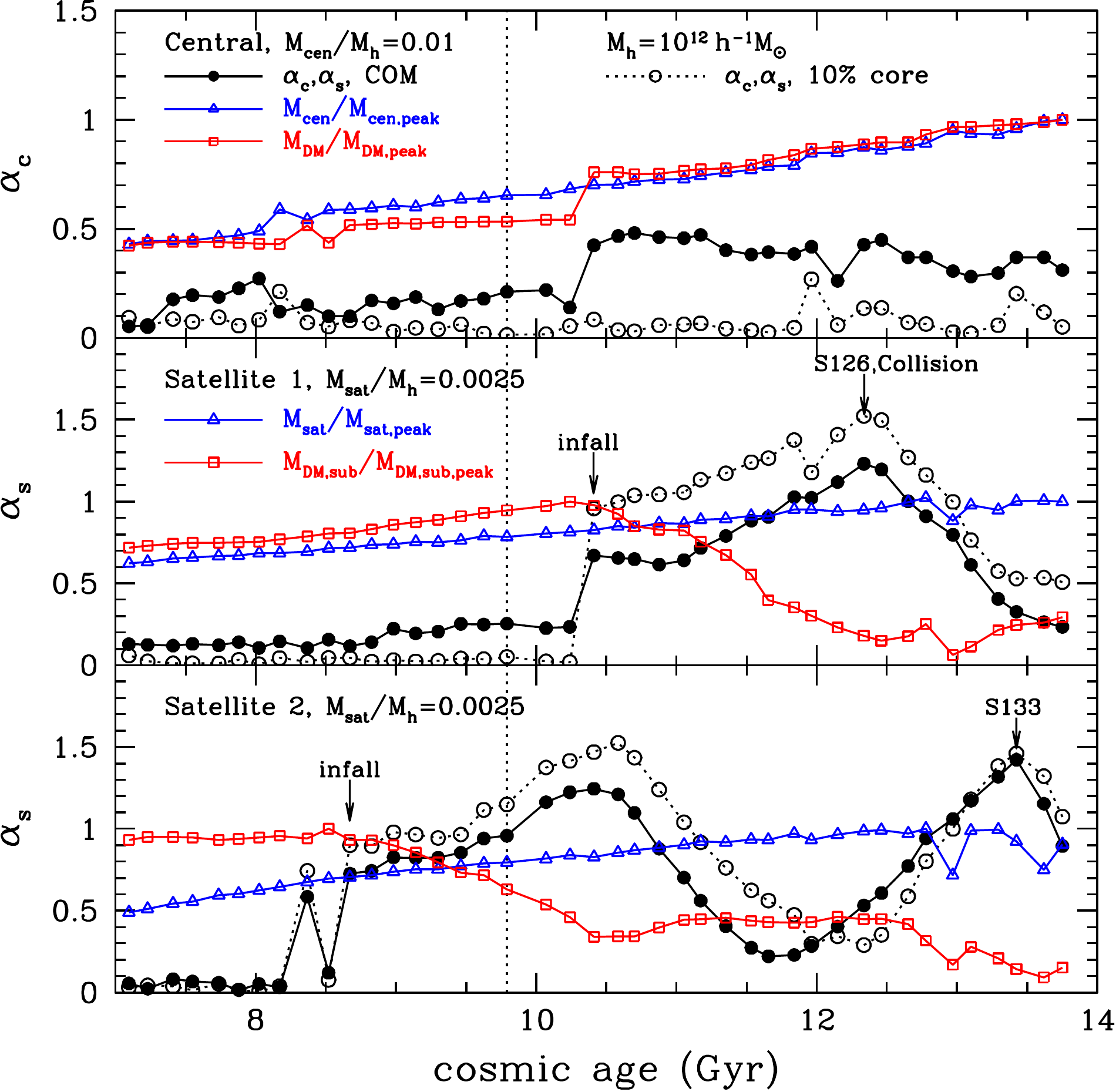}
	\caption{Evolution histories of the galaxy velocity for an example halo system. The top panel shows the evolution of $\alpha_{\rm c}$ of the central galaxy with the cosmic age for the center-of-mass (solid line with filled circles) and $10\%$ core (dotted line with open circles) velocity frames, together with the growth histories of the central galaxy stellar mass (triangles) and the host halo mass (squares). The evolution histories of the galaxy velocity for the two satellite galaxies are shown in the middle and bottom panels. The accretion epochs of these two galaxies onto the host halo are also marked. Before these two satellite galaxies fell into this host halo, they were central galaxies of the progenitor halos. The velocities shown before the infall epoch are actually the values of $\alpha_{\rm c}$ for these two galaxies. The vertical dotted line at $9.79$~Gyrs in each panel indicates the starting epoch for the illustration in Figure~\ref{fig:merger}.
}
	\label{fig:tree}
\end{figure*}
\begin{figure*}
	\epsscale{1.0}
	\plotone{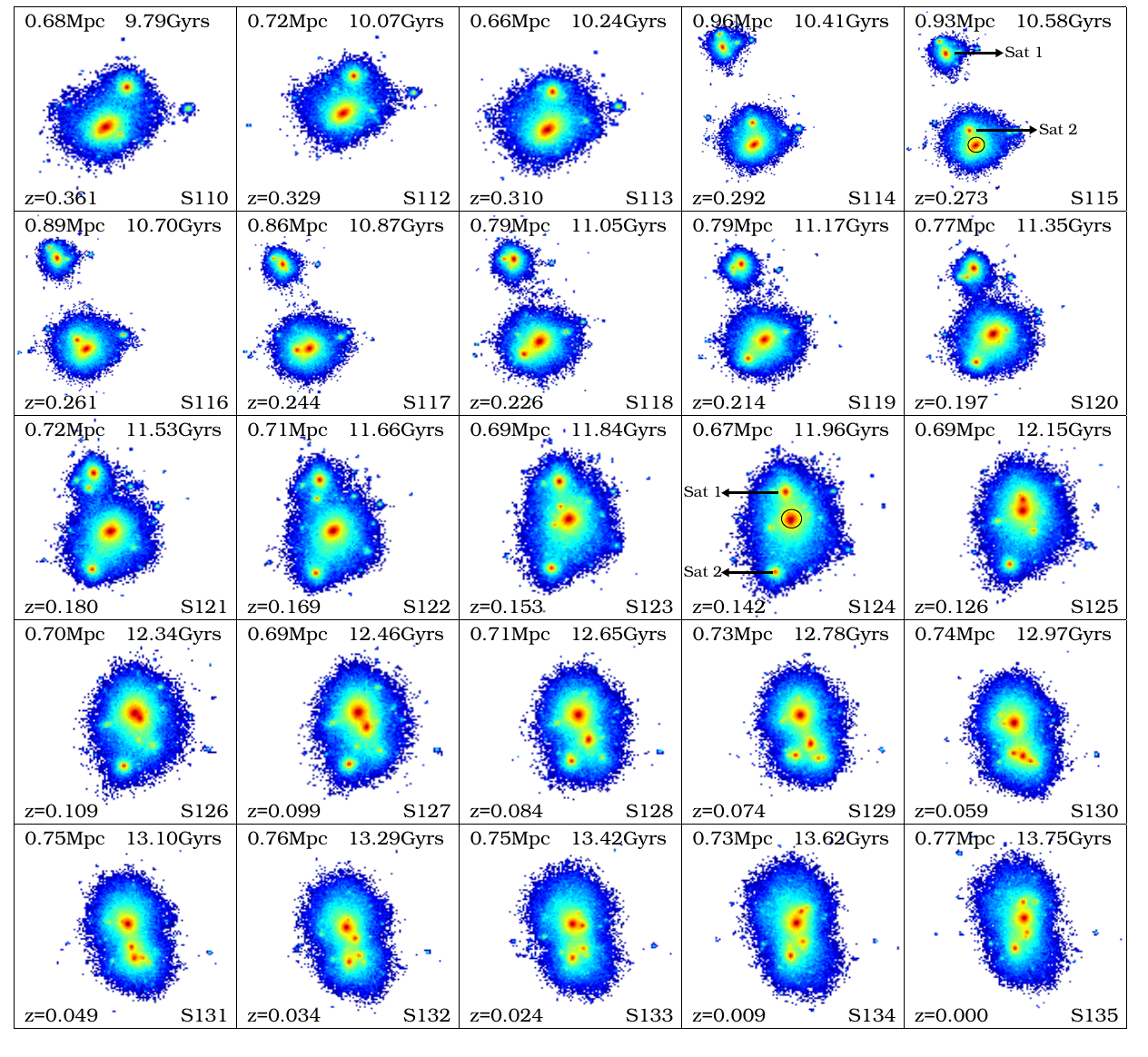}
	\caption{Spatial positions of the particles in the example halo system of in Figure~\ref{fig:tree} starting from the cosmic age of $9.79$~Gyrs ($z=0.361$) to today. The corresponding cosmic age of each panel is shown in the top right, while the physical scale of the panel box, the redshift $z$ and the snapshot number (denoted by the label starting with `S') are indicated in the top left, bottom left and right, respectively. We also show the scale of the 10\% core as black circles in two example snapshots of S115 and S124. The two satellite galaxies in Figure~\ref{fig:tree} are also labeled as `Sat 1' and `Sat 2', respectively.
}
	\label{fig:merger}
\end{figure*}

Since galaxy velocity bias reflects the mutual relaxation of galaxies and their host halos and the tidal and dynamical evolution of galaxies inside the host halos, it would be useful to study the evolution of the veolicity of  central and satellite galaxies in the halo frame. Note that velocity bias is defined in a statistical sense, while here we consider the velocity of each individual galaxy. We still use the notation of $\alpha_c$ and $\alpha_s$, with the definition being $\alpha_{\rm c,s}=|\mathbf{v}_{\rm c,s}-\mathbf{v}_{\rm h}|/\sigma_{\rm v}$ (in parallel to Equation~\ref{eq:velbias}). 

We show an explicit example of the evolution histories of the velocity for a central galaxy and two of its satellite galaxies in Figure~\ref{fig:tree}. This central-satellite system is in a halo of $10^{12}\msun$, the SHMRs for the final epoch ($z=0$) of the central and satellite galaxies are $M_{\rm cen}/M_{\rm h}=0.01$, $M_{\rm sat}/M_{\rm h}=0.0025$ and $M_{\rm sat}/M_{\rm h}=0.0025$, respectively. The two satellite galaxies have similar masses. The top panel shows the evolution of $\alpha_{\rm c}$ of the central galaxy with the cosmic age for the center-of-mass (solid line with filled circles) and $10\%$ core (dotted line with open circles) velocity frames, together with the growth histories of the central galaxy stellar mass (triangles) and the dark matter mass in the host halo (squares). The evolution histories of the galaxy velocity for the two satellite galaxies are shown in the middle and bottom panels, where we also show for comparison the growth histories of the satellite galaxy stellar mass and the dark matter mass of the corresponding subhalos. The accretion epochs of these two galaxies onto the host halo are also marked in the figure. Before these two satellite galaxies fell into this host halo, they were the central galaxies of the progenitor halos. The velocities shown before the infall epoch are actually the values of $\alpha_{\rm c}$ for these two galaxies.

To aid the detailed comparisons, the spatial positions of the particles of the above system are shown in Figure~\ref{fig:merger} starting from the cosmic age of $9.79$~Gyrs ($z=0.361$), which is also indicated by the vertical dotted lines in Figure~\ref{fig:tree}. The corresponding cosmic age is shown in the top right of each panel, while the redshift $z$ and the snapshot number are indicated in the bottom left and right of each panel, respectively.

Comparing Figures~\ref{fig:tree} with \ref{fig:merger}, the central galaxy velocity in the center-of-mass velocity frame is sensitive to the halo mergers, as shown in the sharp increase at the cosmic age of $10.41$~Gyrs (snapshot `S114' in Figure~\ref{fig:merger}) when the satellite 1 was accreted onto the halo. The central galaxy velocity is also sensitive to central galaxy mergers (e.g., at 
cosmic ages of 8.2 Gyrs and 12 Gyrs, when the central galaxy mass has a relatively sharp increase. This was mainly caused by satellite encounters with the 
central galaxy, as shown later). When we switch to the 10\% core velocity frame, the value of $\alpha_{\rm c}$ is less affected by the halo mergers, but has a stronger dependence on galaxy mergers. We note that after the infall of the satellite 1, $\alpha_{\rm c}$ of the central galaxy keeps a high value for quite a long time in the center-of-mass frame because the center-of-mass velocity of the host halo is significantly changed by this merger event. 

In the 10\% core velocity frame, the value of $\alpha_{\rm c}$ only experiences a slight increase at $10.41$~Gyrs (halo merger), and quickly drops to the lower value as in earlier epochs. It means that the halo core velocity is not significantly affected by the mergers. The three spikes in the evolution history of the central galaxy velocity in the 10\% core velocity frame are all related to the direct interaction between the central and satellite galaxies. The two spikes at the cosmic ages of 8.17~Gyrs (snapshot `S100', not displayed in Figure~\ref{fig:merger}) and 11.96~Gyrs (snapshot `S124') are caused by collisions of the satellite galaxies with the central galaxy. The spike of $\alpha_{\rm c}$ at 8.17~Gyrs is not due to the infall of satellite 2, but rather the collision of another satellite galaxy with the central galaxy. We note that the collision of the satellite 1 with the central galaxy actually happens at `S126', while the spike of $\alpha_{\rm c}$ is at `S124'.
This is caused by the fact that the halo core velocity is already affected by the accretion of the satellite galaxy. Even though satellite 1 is still outside the radius of the halo 10\% core, its satellite galaxies (when it was a central) are colliding with the central galaxy at this snapshot. This is due to the fact that the satellite galaxy already enters the radius of the 10\% core at `S124' and the later collision does not change the 10\% core velocity much. The spike at 13.42~Gyrs (`S133') is caused by the change of the central galaxy velocity when the satellite 2 is closest to the central galaxy. 

It is interesting that the evolution history of the satellite galaxy velocity is behaving like a sine curve, due to the oscillations of their trajectories around the halo potential well. The value of $\alpha_{\rm s}$ of a certain satellite galaxy is directly related to its interaction with the central and other satellite galaxies. Therefore, it is hard to model the evolution of $\alpha_{\rm s}$ for a specific satellite galaxy due to the complicated history, but statistically the values of $\alpha_{\rm s}$ can be suitably modeled by taking into account the various aspects and evolution stages in the above sections. 

Although the subhalos of the satellite galaxies suffer mass loss due to the tidal stripping, the stellar masses of the satellite galaxies are steadily growing most of the time. It implies that the velocity field of the satellite galaxies might not be significantly affected by the tidal effects in the halos. The dynamical friction may play a major role in the evolution of the satellite galaxy velocities.

\section{Conclusion and Discussion} \label{sec:conclusion}
In this paper, we use the highest resolution hydrodynamical simulation currently available of the Illustris suite to measure and study the origin of the velocity bias of central and satellite galaxies. For our fiducial model, we use the center-of-mass velocity of the dark matter particles as the halo velocity. We confirm the existence of galaxy velocity bias for both central and satellite galaxies. We find that the values of $\alpha_{\rm c}$ and $\alpha_{\rm s}$ decrease with the ratio between galaxy stellar mass and host halo mass. The dependences on the stellar mass and halo mass for galaxy samples of a given SHMR are weak. The central galaxy velocity bias shows further dependence on the age of the host halo, while the dependence on the galaxy age is very weak for $M_{\rm cen}/M_{\rm h}>10^{-2}$. 

For satellite galaxies, the situation is more complicated. The satellite galaxies with SHMR less than $10^{-3}$ show no dependence on the halo age, while they strongly depend on the galaxy age. The halo and galaxy age dependences for satellite galaxies with SHMR larger than $10^{-3}$ are completely opposite. The velocity for a given satellite galaxy also shows strong dependence on the time after it is accreted onto the halos. Therefore, the satellite galaxy velocity bias also varies with its distance to the halo center. The value of $\alpha_{\rm s}$ and its complex dependence on various parameters reflect the evolution history of the satellite galaxies. As shown in the example of Figures~\ref{fig:tree} and~\ref{fig:merger}, satellite velocity increases in the first 2~Gyrs after accretion, due to the accelerated movement toward the halo centers. It then decreases for another 2~Gyrs due to the dynamical friction and tidal effects in the halo. On average, after about 4~Gyrs since the accretion epoch, the satellite galaxies start the relaxation process in the halos and $\alpha_{\rm s}$ approaches unity with time, i.e., their velocity field gradually becomes consistent with that of the dark matter. 

The exact values of $\alpha_{\rm c}$ and $\alpha_{\rm s}$ are dependent on the reference frame of the halo velocity. When we use the halo velocity defined as the average dark matter velocity within a smaller core radius around the halo center, $\alpha_{\rm c}$ becomes smaller and $\alpha_{\rm s}$ is closer to unity. The central galaxy velocity tends to follow more closely the local dark matter velocity field rather than the bulk velocity of the whole halo. But even when the halo velocity is defined via averaging the dark matter velocities within the central galaxy radius, $\alpha_{\rm c}$ is still non-zero, and approaches a constant value of about $0.04$. This points to the possibility that there is an intrinsic central galaxy velocity dispersion of around $0.04\sigma_{\rm v}$, no matter how the halo velocity is defined. But the dependences of $\alpha_{\rm c}$ on the halo and galaxy ages become much weaker in the $10\%$ halo core velocity frame.

It is shown clearly in Figure~\ref{fig:tree} that in the frame of $10\%$ halo core velocity, the change of the central galaxy velocity bias is mostly caused by the close interaction (e.g., collision) between central and satellite galaxies and the variation of $\alpha_{\rm c}$ is also quickly smoothed out in about $0.5$~Gyrs. Since we assume that the central galaxy and the host halo are located at the same spatial position (potential minimum), a central velocity bias naturally leads to a small offset between the position of the central galaxy and the halo potential minimum, as discussed in detail in \cite{Guo15a}. 

The dependence of the satellite galaxy velocity bias on the halo velocity definition also tends to converge when using a smaller radius around the halo center (Figure~\ref{fig:massratio}). However, the dependence on the SHMR is still significant under different halo velocity definitions. The value of $\alpha_{\rm s}$ is generally less than unity for $M_{\rm sat}/M_{\rm h}>10^{-4}$. Changing the halo velocity from the center-of-mass frame to the $10\%$ core velocity frame does not significantly affect the dependences of $\alpha_{\rm s}$ on the halo and galaxy ages, on the time after the satellite is accreted, or on the distance of the satellite galaxy to the halo center. This is expected from the fact that the value of $\alpha_{\rm s}$ reflects the differences between the velocity dispersion of satellite galaxies and that of the dark matter, and it does not have strong dependence on how the halo velocity is defined.

\begin{figure} %[htbp]
	\epsscale{1.15}
	\plotone{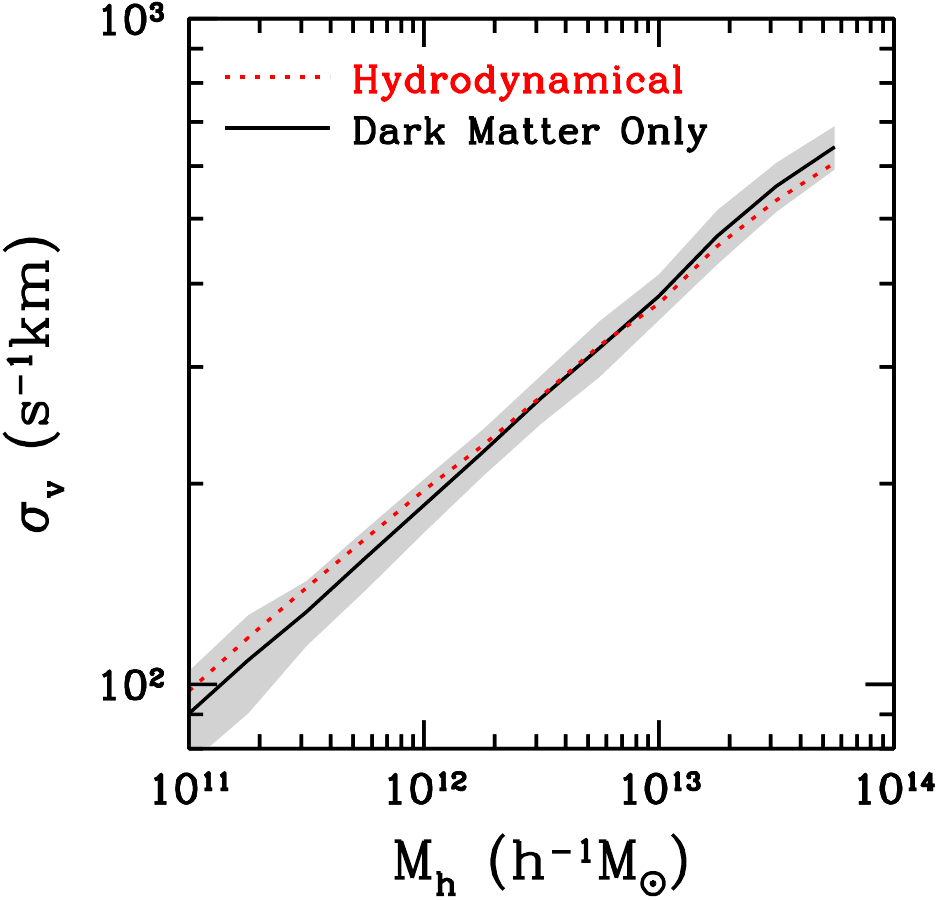}
	\caption{3D dark matter velocity dispersion in halos of different mass $M_{\rm h}$ for the DMO (solid line) and hydrodynamical simulations (dotted line). The shaded region shows the scatter around the velocity dispersion in the DMO simulation.}\label{fig:dm}
\end{figure}
When trying to compare the velocity bias measurements in the hydrodynamical simulation to the result obtained from HOD modeling of observed SDSS galaxy samples \citep{Guo15a,Guo15c}, we need to take into account the impact of baryon that is ignored in the HOD models based on the DMO simulations. We show in Figure~\ref{fig:dm} the 3D velocity dispersions of the dark matter particles in halos of different masses for the DMO (solid line) and hydrodynamical simulations (dotted line). The difference between the velocity dispersions in the two simulations are within $20\%$ for halos of $10^{11}$--$10^{14}\msun$, and they are consistent with each other considering the large scatters of the measurements. Although the halo masses in the full physics run are smaller than those in the DMO run by as much as 25\% for halos of $M_{\rm h}>10^{11}\msun$ in Illustris \citep[see e.g., Fig.~7 of][]{Vogelsberger14a} and other hydrodynamic simulations \citep[e.g.,][]{Sawala13,Velliscig14}, the effect on the dark matter velocity dispersion is only at the level of 9\% (as the velocity dispersion is scaled with halo mass as $\sigma_{\rm v}\propto M_{\rm h}^{1/3}$).

However, since the spatial positions and velocities of the dark matter particles are significantly changed due to baryonic effects, the halo velocities in the hydrodynamical and DMO simulations cannot be directly compared to each other. It is therefore not appropriate to calculate the velocity bias by using the galaxies in the hydrodynamical simulation and the halos in the DMO simulations. For example, although we can match the dark matter halos in the hydrodynamical and DMO simulations by tracing the dark matter particles, the position of a central galaxy in the hydrodynamical simulation can lie far from the center of the corresponding halo in the DMO simulations. But since the velocity dispersion measurements in the two simulations are consistent, the velocity bias measurements based on DMO simulations in the literature are directly comparable to the measurements in the hydrodynamical simulations, once the differences of the cosmological parameters are taken into account. Large DMO simulations are useful particularly for modeling large galaxy survey data. With low-resolution DMO simulations, the center-of-mass halo frame is a natural choice.

\cite{Guo15c} measured the galaxy velocity bias using the SDSS DR7 at $z{\sim}0.1$ for different luminosity threshold samples from faint to luminous galaxies in the center-of-mass halo velocity frame. The average SHMRs for the luminosity threshold samples vary roughly from $10^{-1.9}$ to $10^{-3.5}$ based on the galaxy stellar mass estimated by \citet{Kauffmann03}. However, the SHMRs of central galaxies in the Illustris simulation are much larger than those estimated for the observed galaxy populations \citep{Vogelsberger14a}. For example, the average SHMR for central galaxies in halos of $10^{13}\msun$ is about $10^{-1.9}$ in Illustris-1, but it is around $10^{-2.2}$ from the conditional luminosity function modeling of the DR7 Main galaxy sample in \cite{Yang12} (their Figure~12). This may be partially related to the differences in the definition of halos and the adopted initial mass function of stellar populations.

The galaxy velocity bias is also very sensitive to the SHMR in the center-of-mass halo velocity frame. The direct comparisons of the galaxy velocity bias between the Illustris simulation and the halo modeling results of the observations in the center-of-mass frame thus rely on the accuracy of SHMR in both methods. Therefore, using the $10\%$ halo core velocity provides a better reference frame, as the dependence of galaxy velocity bias on the SHMR is weaker and the appropriate comparisons can be made with accurate SHMR measurements.

The galaxy velocity bias parameters for the SDSS DR7 are measured in the $10\%$ halo core velocity frame in \cite{Guo16}, where $\alpha_{\rm c}$ varies from 0.02 to 0.18 for different luminosity threshold samples (see their Figure~15). More luminous galaxies tend to have larger $\alpha_{\rm c}$, while the SHMR for more luminous galaxies are smaller. This is roughly consistent with the SHMR dependence of $\alpha_{\rm c}$ shown in Figure~\ref{fig:core}. The comparison of the satellite galaxy velocity bias in Illustris-1 with the observations is more difficult, as the $\alpha_{\rm s}$ obtained in \cite{Guo15c} is measured as an ensemble average of a certain luminosity threshold sample and the satellite galaxies with a certain stellar mass can span a large range of halo masses. Nevertheless, the trend of $\alpha_{\rm s}$ with the SHMR shown in the right panel of Figure~\ref{fig:core} is consistent with \cite{Guo15c}. 

For the theoretical modeling of the galaxy velocity bias in the halo framework, using the $10\%$ halo core velocity is seemingly a better choice because it is relatively insensitive to the halo merger events. Accurate modeling of $\alpha_{\rm c}$ still requires to carefully take into account the SHMR dependence, which can be done by selecting galaxies in various stellar mass ranges and adopting a power-law relation between $\alpha_{\rm c}$ and the SHMR. The modeling of $\alpha_{\rm s}$ is still challenging, especially to consider the dependence on the time after accretion and the radial distribution of satellite galaxies. A better understanding of the origin and properties of velocity bias will help the intepretation of small-scale reshift-space distortion in galaxy clustering and the use of clustering data to learn about the kinematic evolution of galaxies.

In this paper, we investigate the dependence of the galaxy velocity bias on the intrinsic galaxy properties, such as the stellar mass and galaxy age. However, in observation, these properties are not directly measured, and their inferences depend on fitting the spectra or spectral energy distribution of galaxies with stellar population synthesis models. While studying the dependence of galaxy velocity bias on luminosity or color would be more relevant for the comparison with that inferred from observation \citep[e.g.,][]{Guo15c}, the Illustris simulation does not exactly reproduce the observed galaxy luminosity function and the bimodal distribution of galaxy color \citep[see Figs.~10 and~14 of][]{Vogelsberger14a}. In addition, dust effects are not included in the predicted luminosity from the Illustris simulation. Therefore, a direct comparison to the luminosity dependence of galaxy velocity bias measured in \cite{Guo15c} is not straightforward, and for the theoretical investigation we focus on the dependences of galaxy velocity bias on galaxy intrinsic properties.

The halos used in the model of \cite{Guo15c} are identified through the \texttt{ROCKSTAR} phase-space halo finder \citep{Behroozi13}, while the halo catalog in Illustris used in this study is constructed using the FOF and \texttt{SUBFIND} algorithm based on the galaxy spatial distribution. With phase-space information \texttt{ROCKSTAR} algorithm removes unbound particles in each halo, while FoF and \texttt{SUBFIND} do not. As less than 2\% of halo particles are found to be unbound \citep{Behroozi13}, whether to remove unbound particles is unlikely to affect the amplitudes of velocity bias. The tests with FoF and spherical oversensity (SO) halos in \citet{Guo15a} show that the central galaxy velocity bias is not affected by the halo definition and that FoF halos lead to a lower amplitude in satellite velocity bias (but still consistent with that from SO halos). The amplitudes of the galaxy velocity bias we infer from Illustris is in broad agreement with that constrained in \cite{Guo15c}, supporting that different halo definitions/finders do not make significant differences in velocity bias.

Although our study of the galaxy velocity bias is based on the state-of-art hydrodynamical simulations of the Illustris suite, they still have their own limitations as in any other hydrodynamical simulations, such as the over-predicted star formation rate at $z<1$ \citep{Vogelsberger14a, Genel14}, the inefficient radio-mode AGN feedback for massive halos, and the two times larger sizes for low-mass galaxies of $\sim10^{10}\msun$ \citep{Snyder15}. These discrepancies are related to the baryonic effects in the adopted galaxy formation models. But as suggested by the findings of \cite{Hellwing16}, the baryonic effects would only have minimal influence on the galaxy and halo velocity fields and we expect our results on the galaxy velocity bias to be less dependent on the way baryonic physics is implemented. However, given the complicated evolution histories of the satellite galaxies, the dependences of the satellite velocity bias on various galaxy properties might potentially be more influenced by the details in the galaxy formation models. With more realistic hydrodynamical simulations in future, we will be able to have better understanding of the evolution of the galaxy velocity field.

\section*{Acknowledgments}
This work is supported by the 973 Program (No. 2015CB857003) and NSFC-11543003. HG is supported by the 100 Talents Program of the Chinese Academy of Sciences. ZZ was partially supported by NSF grant AST-1208891. IZ acknowledges support by NSF grant AST-1612085 and by a CWRU Faculty Seed Grant.

\end{document}